\documentclass[twocolumn,prl,preprintnumbers,amsmath,amssymb]{revtex4}

\usepackage{xspace}
\usepackage{graphicx}
\usepackage{dcolumn}
\usepackage{bm}

\def\beq{\begin{equation}}
\def\eeq{\end{equation}}

\def\bea{\begin{eqnarray}}
\def\eea{\end{eqnarray}}

\newcommand{\GeV}{\ensuremath{\mathrm{~GeV}}}

\newcommand{\pb}{\,{\rm pb}}
\newcommand{\fb}{\,{\rm fb}}

\newcommand{\Fig}[1]{Fig.~\ref{#1}}

\def\Zp{Z^\prime}
\def\Wp{W^\prime}
\def\Afb{A_{FB}}
\def\Afbp{A_{FB}^{+}}
\def\Afbm{A_{FB}^{-}}
\def\mtt{m_{t\bar{t}}}
\def\Vp{V^\prime}
\def\cth{\cos \theta}
\def\MET{E_T^{\textrm{miss}} }
\def\={\,=\,}

\begin{document}

\preprint{MCTP-11-28}
\preprint{CERN-PH-TH/2011-198}

\title{Top asymmetry and the search for a light \\
hadronic resonance in association with single top}

\author{Sunghoon Jung}\email{jungsung@umich.edu}
\author{Aaron Pierce}\email{atpierce@umich.edu}
\affiliation{Michigan Center for Theoretical Physics, Department of Physics,
University of Michigan, Ann Arbor, MI 48109}

\author{James D. Wells}\email{jwells@umich.edu}
\affiliation{Michigan Center for Theoretical Physics, Department of Physics,
University of Michigan, Ann Arbor, MI 48109}
\affiliation{CERN Theoretical Physics (PH-TH), CH-1211 Geneva 23, Switzerland}

\date{\today}

\begin{abstract}
The exchange of a light $t$-channel flavor-changing gauge boson, $\Vp$, with mass $\sim m_{top}$ remains a leading explanation for the anomalous forward backward asymmetry in top quark production at the Tevatron. Unlike other ideas, including heavier $t$-channel mediators, the light $\Vp$ model is not easily seen in the $\mtt$ distribution.  We advocate a more promising strategy. While current analyses at hadron colliders may not be sensitive, we propose searching for a  $jj$ resonance in association with single top that may allow discovery in existing data. Deviations in the lepton charge asymmetry in this sample should also be present.
\end{abstract}
\maketitle

{\it Top asymmetry persists.} A tantalizing anomaly persists in the measurement of the forward-backward asymmetry of top quark at the Tevatron. Three independent measurements of $\Afb$ have been carried out in the $t\bar{t}$ rest frame that all yield large values. Two are from $\ell +j$ channel \cite{cdf:afb9724,Abazov:2011rq}:
\bea
 \Afb &=& \, 19.6 \pm 6.5 \% \, (\textrm{D0, 5.4} \fb^{-1}),\\
  \Afb &=&\,15.8 \pm 7.4 \%\, (\textrm{CDF, 5.3} \fb^{-1}),
\eea
while the other is from $\ell \ell$ channel utilizing precise measurement of lepton momenta \cite{cdf:10436}
\beq
 \Afb^{\ell \ell}= 42.0 \pm 15.8 \%\, (\textrm{CDF, 5.1} \fb^{-1}).
\eeq
These independent results are all $\sim 2 \sigma$ away from NLO SM predictions $\Afb = 5.8 \pm 0.9\, (\ell+j),\, 6.0 \pm 1.0 \%\, (\ell \ell)$ \cite{cdf:afb9724,cdf:10436,Kuhn:1998jr,Kuhn:1998kw,Bowen:2005ap,Ahrens:2010zv}. The CDF experiment also sees evidence for a particularly large value of $\Afb$ for $\mtt > 450 $ GeV \cite{Aaltonen:2011kc}, the data from D0 \cite{Abazov:2011rq} do not show such a pronounced rise but are consistent with a more modest increase.

\bigskip
{\it New physics explanation.} A new flavor-changing $t$-channel mediator can explain the elevated $\Afb$ measurement. Such a model with a gauge boson $\Vp$ with mass $m_{\Vp} \sim m_{top}$ and $\Vp$--$u$--$t$ coupling was proposed and studied in ref. \cite{Jung:2009jz}. Unlike this original Abelian gauge model, non-Abelian versions \cite{Jung:2009jz,Jung:2011zv,Cheung:2009ch,Barger:2010mw} can simultaneously explain the absence of same-sign dilepton events (or same-sign tops) at either the Tevatron \cite{cdf:10466} or the LHC \cite{Chatrchyan:2011dk,Collaboration:2011vj}. The light $\Vp$ is also in the proper mass range \cite{Jung:2011ua,Buckley:2011vc,Nelson:2011us,Jung:2011ue,Ko:2011vd} to give contributions to $Wjj$ excess seen at CDF \cite{Aaltonen:2011mk}. However, it is difficult for the flavor changing couplings of these models to fully explain the excess \cite{Jung:2011ua}. Conversely, these models will not be in conflict with the data, even if the full $Wjj$ excess does not persist in future measurements, as indicated by the recent D0 result \cite{Abazov:2011af}.

While the models involving $t$-channel exchange of heavier exotics (mass of several hundred GeV or more) have been studied in great detail, see, e.g. \cite{Cheung:2009ch, Shu:2009xf, Arhrib:2009hu, Dorsner:2009mq, Jung:2009pi, Barger:2010mw, Cao:2010zb, Xiao:2010hm, Jung:2010yn, Cheung:2011qa, Shelton:2011hq, Bhattacherjee:2011nr, Barger:2011ih, Gresham:2011dg, Patel:2011eh, Grinstein:2011yv, Barreto:2011au, Ligeti:2011vt, AguilarSaavedra:2011vw, Gresham:2011pa,Krohn:2011tw,Shu:2011au,Gresham:2011fx}, the LHC consequences of a light $t$-channel mediator ($m_{\Vp} \sim m_{top}$) remains relatively unexplored. Although the light and heavy $\Vp$ share the property that a large $\Afb$ can be easily generated, a light $\Vp$ has potentially drastically different collider phenomenology.  In particular, if $m_{\Vp} \lesssim m_{t}$, phase space suppression means that even a small diagonal coupling to light quarks leads to the dominant decay mode $\Vp \to jj$. In this letter, we compare and contrast the light $\Vp$ model with the heavier $t$-channel $\Vp$ models, and suggest that the recent ATLAS measurement of $\mtt$ \cite{ATLAS:2011-087} may already favor a light mediator. We discuss the inadequacies for testing this model by using this distribution, and we present alternate search strategies utilizing the single top data sample.

Our benchmark model descends from a non-Abelian $SU(2)_X$ horizontal symmetry \cite{Jung:2011zv} where $(u \, t)_R$ form a doublet. There are new states with dominantly flavor off- diagonal couplings, which we call $\Wp$, and a new state with dominantly flavor preserving couplings, which we call $\Zp$.  The Tevatron anomaly is explained dominantly via the $\Wp$.  The parameters of the model are: $M_{\Wp}=160\GeV,\,$ $M_{\Zp}=80\GeV,\,$ $\alpha_X=0.045,\,$ $\cth=0.995$.   Here $\theta \neq 0$ represents a very small mismatch between the quark mass eigenstates and the eigenstates of $SU(2)_X$ that allows for $\Wp$ to decay to $u\bar{u}$.  We call this model point as ``Model A". This model point is very similar to Model A considered in \cite{Jung:2011ua} as well as the best point model of \cite{Jung:2009jz}. Predictions of the $\Afb$ and top production cross sections are in good agreement with present data.
A summary of the Tevatron $A_{FB}$ predictions are presented in Table \ref{tab:afbresults}.
A closely related observable that can be measured at the LHC is $A_{boost}$ \cite{Jung:2011zv}. This is defined to be the top asymmetry with respect to the $t\bar t$ boost direction. After cuts but before unfolding we predict $2.5\%$. CMS measures $-0.7\%$ with an unknown error that is not expected to be greater than $3.8\%$ \cite{cms:11-014}. This measurement does not appear to  constrain the theory at present. We provide through supplementary notes~\cite{oursuppl} more discussion of these points.
We use this model throughout this paper to discuss the physics of a light $\Vp$. However, our results should be broadly applicable to a large class of models, e.g. left-right asymmetric $\Wp$ model \cite{Cheung:2009ch,Barger:2010mw,Barger:2011ih,Cheung:2011qa}, or a $t$-channel scalar mediator \cite{Shu:2009xf,Arhrib:2009hu,Dorsner:2009mq,Cui:2011xy}.  The crucial ingredient is a light mediator with small  coupling to light quark pairs, in addition to the larger couplings to $u/d$--$t$ that explain the $\Afb$ result.
\begin{table*}
\begin{tabular}{@{\hspace{0.2cm}} c @{\hspace{0.2cm}} c @{\hspace{0.2cm}} c @{\hspace{0.2cm}}c @{\hspace{0.2cm}} }
\hline \hline
 & $\Afb$ & $\Afbp$ & $\Afbm$  \\
\hline
Model A & $19\%$  & $35 \to 21\%$ & $5 \to 6\%$   \\
\hline
CDF \cite{cdf:afb9724,Aaltonen:2011kc} & $15.8 \pm 7.4 \%$ & $47.5 \to 26.6 \pm6.2 \%$  &  $-11.6 \to -2.2 \pm4.3\%$ \\
D0 \cite{Abazov:2011rq} & $19.6 \pm  6.5\%$                  & - $\to 11.5 \pm 6.0 \%$   &  - $\to 7.8 \pm 4.8 \%$   \\
\hline
SM  & $5.8 \pm 0.9 \%$ & $8.8 \to 4.3 \pm 1.3 \%$ & $4.0 \to 1.3 \pm 0.6 \%$ \\
\hline \hline

\end{tabular}
\caption{List of various Tevatron asymmetry results. $\Afbp$($\Afbm$) is defined for $\mtt >(<) 450 \GeV$. All model predictions are before cuts except for the $A_{FB}^{+,-}$ results shown after arrows. These are obtained with selection cuts and bin-to-bin migration effects (see ref.\cite{Jung:2011zv} for more detail), thus can be compared with reconstruction level results (data-background level) of CDF and D0 shown. See text for discussion of closely related LHC observable.} \label{tab:afbresults} \end{table*}

\begin{figure}
\includegraphics[width=0.5\textwidth]{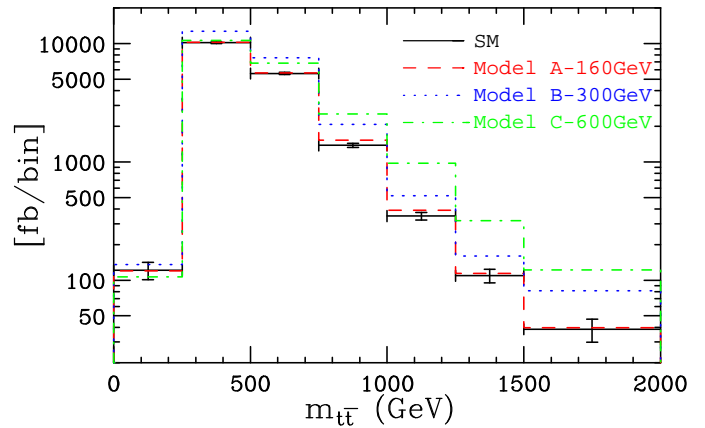}
\caption{$\mtt$ distributions for SM(black solid), Model A(red dash), Model B(blue dot), Model C(green dotdash) at the LHC7. Simulated distributions are shown after applying the ATLAS dRmin algorithm. Shown error bars correspond to MC + $1\fb^{-1}$ statistical uncertainty, but systematic uncertainty is currently larger \cite{ATLAS:2011-087}. Model B is contrasted to show large effects of $gu \to tV' \to t\bar{t}q$, and Model C illustrates the relatively poor acceptance of Model A at high-$\mtt$ bin.}
\label{fig:mtt-all}
\end{figure}

\bigskip
{\it Relevance of $m_{t\bar t}$?}
We now discuss how the recent ATLAS measurement of $\mtt$ \cite{ATLAS:2011-087} is favorable to a $\Vp$ model with a light $t$-channel mediator. To this end, we contrast our model with two heavier $\Vp$ models.  We call these models ``Model B" with $M_{\Wp}=300 \GeV,\, \alpha_X=0.12$ and ``Model C" with $M_{\Wp}=600\GeV,\, \alpha_X=0.38$. The coupling constants are chosen to produce an identical $\Afb \simeq 19\%$.  For simplicity, in these two models we assume that the $SU(2)_{X}$-neutral $\Zp$ is  sufficiently heavy that it has decoupled. In \Fig{fig:mtt-all}, we show $\mtt$ distributions for these models. Event samples are obtained by MadGraph \cite{Alwall:2007st} interfaced with Pythia \cite{Sjostrand:2003wg} (MLM matched \cite{Mangano:2001xp,Alwall:2007fs} with up to one extra jet)  and PGS detector simulation (with an anti-$k_T$ jet algorithm implemented by ourselves).
Finally, predictions are obtained by employing the ATLAS dRmin $\mtt$ reconstruction algorithm \cite{ATLAS:2011-087}.

The LHC data is consistent with the SM $m_{t\bar t}$ distribution. When comparing the new physics models in \Fig{fig:mtt-all}, it is clear that Model A is most similar to the SM result and thus consistent with the data.   This agreement comes about in a non-trivial way as we will discuss below.  Model B suffers from sizable contributions from the process  $gu\to t\Vp \to t\bar{t}j$. This contribution is not only large ($\sim 20\pb$), but also has a different $\mtt$ distribution than the true SM $t\bar{t}$.  This contribution shows up as an excess in every bin; in fact, this model likely yields a too large total $\sigma_{t \bar{t}}$. Contributions of this type are absent for Model A because the 160 GeV $\Vp$ dominantly ($\gtrsim$ 95\%) decays to $jj$, and so this process does not enter the $t \bar{t}$ sample.  The $\Vp$ of Model C is sufficiently heavy that, happily, similar processes do not contribute to the $\mtt$ sample.  However, the heaviness of the mediator regulates the $t$-channel (Rutherford) enhancement. As a result, top quarks from Model C are not produced too far in the forward region, and they have relatively large acceptance, leading to a large deviation in the $m_{t \bar{t}}$ distribution.  This is to be contrasted with Model A which produces very forward top quarks as a result of  a stronger Rutherford enhancement. Since the tops are so far forward, the acceptance is drastically reduced \cite{Jung:2011zv,Gresham:2011pa}, and agreement with the data is better than one might anticipate.  Additionally, our simulation shows that the reconstruction algorithms used by ATLAS, CMS, CDF spread out true $\mtt$ distributions in such a way that events one thinks should fall into low-$\mtt$ bins actually fall into the higher $\mtt$ bins (we refer to our supplementary notes~\cite{oursuppl} for more figures with details). This contamination in the upper bins can dominate over the true high $m_{t\bar t}$ contributions from new physics, diluting the sensitivity.   In summary, at present Model A seems completely consistent with the data.

One possible way to better isolate the Model A contribution would be to use a $\chi^2$ method (see e.g., refs.\cite{cdf:afb9724,cms:10-007}) where a maximum cut on $\chi^2$ is employed on a completely reconstructed $t \bar{t}$ event.  However, even employing this method, we deem it unlikely that $\mtt$ would be an optimal discovery mode
for this model.  For more promising approaches, we turn to the single top sample.

\bigskip
{\it Concomitant resonance.} There is abundant production of the $\Vp$ in association with a single top quark in $gu \to t \Vp \to t jj$. The signal event topology is $W+3j$(with one $b$-tag).
Before discussing how this sample can yield a discovery, we first assert that current analyses at hadron colliders would not see the model.
One might think that cuts that isolate single-top should be efficient for this model because the signal cross section is $\sigma(t\Vp) \sim 1 (60)\pb$ at the Tevatron (LHC7), and event topology is similar to SM single-top production. However, most of cut-based single top analysis have been optimized in $W+2j$ exclusive channel so far \cite{ATLAS:2011-027,cms:10-008} where our model's contribution is small \cite{Jung:2011ua}, and these measurements suffer from a sizable systematic uncertainty. One exception is from recent ATLAS note \cite{ATLAS:2011-101}, and will be discussed later with $A_C^\ell$.
Also, it has been suggested \cite{Craig:2011an} that the tail of the $H_T(j)$ distribution in the single top sample is a sensitive probe of new physics contributing to $\Afb$.  However, as $\Vp$ is light in our case, the contribution from $t\Vp$ process does not surpass the $t\bar{t}$ background contributions, and thus remains hidden. On the other hand, D0 has data in the $W+3j$ exclusive channel resulting from a search for $Wh$ using $m_{jj}$ (with one extra jet radiated) \cite{Abazov:2010hn}. However, the $m_{jj}$ in this analysis is reconstructed using any two leading jets (tagged or not) while $\Vp$ decays to (untagged) light jets. This dilutes the signal.  We conclude that at present, this model is not ruled out.

\begin{figure}
\includegraphics[width=0.5\textwidth]{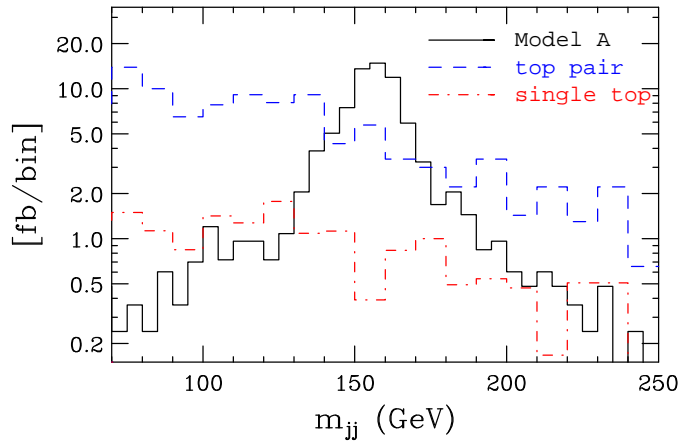}
\caption{$m_{jj}$ distribution at the LHC7 after all discovery cuts described in text. In addition to Model A signal, dominant background $t\bar{t}$ as well as SM single top contributions are shown.}
\label{fig:mjj-lhc7}
\end{figure}

\bigskip
{\it Resonance at LHC.} It appears possible to reconstruct the $\Vp$ resonance in the sample where it is produced in association with a single top through $gu \to tV' \to tjj$. The event topology that we seek for $\Vp$ resonance is
\begin{itemize}
\item Three jets exclusive final state. Amongst these three, we require one $b$-tag. The two untagged jets are used to construct $m_{jj}$.
\item One and only one charged lepton (either $e$ or $\mu$).
\item Missing energy $\MET$.
\label{item:topology} \end{itemize}
Quantitatively, inspired by the  ATLAS single top analysis \cite{ATLAS:2011-101}, we initially apply the following basic kinematic selection cuts (set A):
\begin{itemize}
\item jet: $p_T>25 \GeV$, $\eta < 4.5$
\item lepton: $p_T>25 \GeV$, $\eta <2.5$
\item $\MET>25$ GeV, $M_T^W(\ell,\nu) > 60 \GeV- \MET$
\end{itemize}
These basic cuts are insufficient to reveal the $\Vp$ resonance due to backgrounds of $t\bar{t}$ and (sub-dominantly) $W+j$ \footnote{$W+j$ are generated with Alpgen \cite{Mangano:2002ea} at parton-level.}. To enhance the signal, we propose an additional set of hard cuts based on our MC to extract the resonance signal (set B):
\begin{itemize}
\item $135 \leq m_{jj} \leq 175 \GeV$
\item $\Delta R (j_1, j_2) < \pi$
\item $p_T$(lead $j ) > 90 \GeV$
\item $H_T(j) > 200 \GeV$
\end{itemize}
The cuts are applied to untagged jets, and $H_T(j)$ is the scalar sum of the $p_T$ of all three jets (tagged or not). After all these cuts, the $m_{jj}$ distribution looks like \Fig{fig:mjj-lhc7}. Significance of the resonance signal can be estimated as in Table \ref{tab:mjjresult}. Systematic uncertainty of the single top sample could be significant.
If systematics are brought under control and the statistical uncertainty dominates  a 5$\sigma$ observation may already be possible in $1\, {\rm fb}^{-1}$ of LHC7 data. Thus, current data may be sufficient to observe a $\Vp$ resonance once optimal cuts are applied.  Alternately, very strong bounds can be placed on models where the $\Vp$ dominantly decays to a pair of jets.

\begin{table}
\begin{tabular}{@{\hspace{0.2cm}} c @{\hspace{0.4cm}} c @{\hspace{0.3cm}}}
\hline \hline
backgrounds  & $\sigma$ after discovery cuts\\
\hline
$t\bar{t}$ & 0.20 pb \\
Single top ($t$-channel) & 0.019 pb \\
Single top ($tW$) & 0.016 pb \\
$W+j$ & 0.080 pb \\
$Wb\bar{b}$ & 0.012 pb \\
\hline
Model A & 0.33 pb \\
\hline
$S/\sqrt{B}$ & $5.7 \sqrt{ {\cal L} / 100\pb^{-1} }$ \\
\hline \hline

\end{tabular}
\caption{$V'$ resonance search result at the LHC7 after all discovery cuts (Set B) described in text.}
\label{tab:mjjresult} \end{table}

\bigskip
{\it Resonance at Tevatron.} We now discuss discovery prospects of the resonance at the Tevatron. Based on the Tevatron single top analysis \cite{Abazov:2009ii,Aaltonen:2010jr}, we apply the following discovery cuts (masses in GeV):
\begin{itemize}
  \item Three jets with $p_T > 25 \GeV,\, p_T($lead $j)>50 \GeV,\, \eta <2.8$.
  \item One $b$-tagged jet. Two untagged jets for $m_{jj}$.
  \item One lepton ($e$ or $\mu$) with $p_T > 20 \GeV,\, \eta <1.6$.
  \item $\MET > 25 \GeV$. $M_T^W(\ell,\nu) \geq 10 \GeV$.
  \item $H_T(all) \geq 220 \GeV$.
\end{itemize}
where $H_T(all)$ is the scalar sum of the $p_T$ values of all three jets, transverse missing energy and leptons. Then we count the number of events within
\beq
125 \GeV \leq m_{jj} \leq 165 \GeV.
\eeq
After all cuts, $t\bar{t}$ remains the dominant background. Although $S/B = 0.35$ is small, the statistical significance can be substantial $S/\sqrt{B} = 2.0 \sqrt{{\cal L} / 1\fb^{-1}}$.  Systematic errors may be important, but
prospects for an observation of the resonance at the Tevatron appear promising.


\bigskip
{\it Single lepton charge asymmetry.} A complimentary observable that could confirm the existence of light $\Vp$ is the single lepton charge asymmetry~\cite{Bowen:2005xq}.
This observable is defined using the well-measured sign of single lepton as
\beq
A_C^{\ell} \, \equiv \, \frac{ N(\ell^+ X) - N(\ell^- X)}{N(\ell^+ X) + N(\ell^-  X)}.
\eeq
A signal for these observable in our model arises from the $gu \to t\Vp \to tjj$ process (this observable has also been studied for different processes \cite{Craig:2011an,Rajaraman:2011rw}). Valence $u$ quarks at the LHC lead to an asymmetry in the charge of a $t$ (and hence lepton) in the final state. After applying the basic kinematic cuts (set A), we estimate $A_C \sim 75\%$ for this signal process. Different SM processes also give non-zero $A_C^\ell$ as tabulated in Table \ref{tab:ac-lhc7} \footnote{One of the dominant backgrounds, $t\bar{t}$, does not exhibit $A_C^{\ell}$ at LO. The NLO correction does indicate that $\bar{t}$ will be produced more centrally than $t$, with the effect of producing a small negative asymmetry when selection cuts are made. However, we will ignore this small effect in the following analysis.}. The values of $A_C^{\ell}$ in this Table were generated with the use of our Monte Carlo event samples. Adding all these contributions weighted properly by individual rate (from ATLAS single top analysis \cite{ATLAS:2011-101}), we predict $A_C^\ell({\rm SM}) = 0.10 \pm 0.014$(stat), and $A_C^\ell({\rm Model\, A}) = 0.19 \pm 0.013$(stat) if the new physics contribution is also added. While these values are very promising, we emphasize that the errors quoted are only statistical.  Understanding systematic errors and their correlation between the $N(\ell^+)$ and $N(\ell^-)$ may play an important role.  We illustrate this point through a brief discussion of the current experimental situation.

A naive combination of ATLAS data in a strongly enriched single-top sample for the 3-jet exclusive state (see Table 2 of  ref.\cite{ATLAS:2011-101}) gives, $A_{C}^{\ell}$(hard cuts) $= 0.10 \pm 0.10$(stat).  The larger statistical error results from harder cuts than we considered above. This value should be compared with theory simulation results for $A_{C}^{\ell}$ applying similar cuts. We find $A_C^{\ell}$(SM, hard cuts) $= 0.18$, and by adding our new physics contribution $A_C^{\ell}$(Model A, hard cuts) $= 0.29$.  Naively, the data favors the SM.  However, depending on correlations, the systematic errors quoted in Table 2 of \cite{ATLAS:2011-101}, could easily yield $\delta A_{C}^{syst} = 0.1$ or more. The potential presence of such a large systematic error precludes at present any defensible statement regarding the model's compatibility with the data.  Nevertheless, $A_C^{\ell}$ seems a promising observable, and a dedicated $A_C^\ell$ analysis of present data with special attention to systematics may be sufficient to draw a conclusion.

\begin{table}[t]
\begin{tabular}{ @{\hspace{0.2cm}} c @{\hspace{0.4cm}} c @{\hspace{0.3cm}} c @{\hspace{0.3cm}} }
\hline\hline
backgrounds & ATLAS total rate & $A_C^\ell$ \\
\hline
$t\bar{t}$  & $1847$ events & 0\\
$W + j$     & $1930$ events & 0.2\\
Single top  & $385$ events & 0.3\\
others      & $668$ events & 0\\
\hline
$t \Vp$ (Model A)  & $780$ events & 0.75 \\
\hline
Total (SM only) & $4830$ events & $0.10 \pm 0.014$(stat)\\
Total (Model A) & $5610$ events & $0.19 \pm 0.013$(stat)\\
\hline
\hline
\end{tabular}
\caption{Predicted background total rates and $A_C^{\ell}$ in the $W+3j$(1 b-tag) topology defined by cut Set A. Predicted ATLAS rates with $0.7 \fb^{-1}$ of data are from Table 1 of ref. \cite{ATLAS:2011-101}. A predicted statistical error is shown. $A_C^\ell$ is obtained by using our MC samples.} \label{tab:ac-lhc7}
\end{table}

\bigskip
{\it Cross-check advocacy.} The persistence of the $\Afb$ anomaly begs for a cross check.  We have argued that a search for a $jj$ resonance in association with a top quark is a definitive signal for a light $t$-channel $\Vp$.  In time, $A_C^{\ell}$ may also prove to be a useful cross-check. These searches, and carrying out the suggested analysis techniques described above, may serve to conclusively discover or refute this model.


\bigskip
\emph{Acknowledgements:} We thank M. Gresham and I.-W. Kim for useful discussions.  This work is supported by DOE Grant \#DE-FG02-95ER-40899. The work of AP is also supported in part by NSF Career Grant NSF-PHY-0743315.

\bibliography{jjsingletop}
\bibliographystyle{apsrev}

\end{document}